\documentclass[preprint2]{aastex}

\usepackage[dvips]{epsfig}
\usepackage[dvips]{rotating}

\bibpunct{(}{)}{;}{a}{}{,}

\newcommand{\he}{HE~0107$-$5240}
\newcommand{\cd}{CD~$-38^{\circ}\,245$}

\newcommand{\vrad}{$v_{\mbox{\scriptsize rad}}$}

\slugcomment{}

\shorttitle{Oxygen abundance of {\he}}
\shortauthors{Bessell et al.}

\begin{document}

\title{On the oxygen abundance of {\he}\altaffilmark{1}}

\author{M. S. Bessell\altaffilmark{2}, N. Christlieb\altaffilmark{2,3}, and
  B. Gustafsson\altaffilmark{4}}

\altaffiltext{1}{Based on observations collected at the European Southern
  Observatory, Paranal, Chile (Proposal Number 70.D-0009).}
\altaffiltext{2}{Research School of Astronomy \& Astrophysics, Mount Stromlo
    Observatory, Cotter Road, Weston, ACT 2611, Australia}
\altaffiltext{3}{Hamburger Sternwarte, Gojenbergsweg 112, D-21029 Hamburg, Germany}
\altaffiltext{4}{Uppsala Astronomical Observatory, Box 524, SE-75239 Uppsala, Sweden}

\begin{abstract} 
  We have determined the oxygen abundance of {\he} from UV-OH lines detected in
  VLT/UVES spectra. Using a plane-parallel LTE model atmosphere, we derive
  $\mbox{[O/Fe]}= +2.4$, and a similar analysis of {\cd} yields $\mbox{[O/Fe]}=
  +1.0$. We estimate systematic errors due to 3D effects to be in the order of
  $0.3$ to $0.4$\,dex. That is, our derived O abundances are likely
  overestimates: effects from thermal inhomogeneities due to convection may
  require that the abundances should be reduced by $0.3\mbox{--}0.4$\,dex or
  even more. Radial velocity data for {\he} based on high-resolution spectra
  show that over a time span of 373 days the radial velocity was constant at
  $44.5$\,km/s, with a $1\,\sigma$ scatter of the measurements of
  0.5\,km/s. However, it can not yet be ruled out that {\he} is a very long
  period and/or low amplitude binary. These results provide new constraints on
  scenarios for the origin of the abundance pattern of {\he}. In particular, it
  seems unlikely that the large overabundances of CNO have been produced in a
  medium-mass AGB star which later evolved to a white dwarf. The oxygen
  abundance of {\he} is significantly smaller than the prediction of Umeda \&
  Nomoto (2003) from calculated yields of a $\sim 25\,M_{\odot}$ Population~III
  star exploding as a supernova of low explosion energy ($E_{\mbox{\scriptsize
  exp}}=3\cdot 10^{50}\,\mbox{erg}$) with mixing and fallback. The scenario of
  Limongi et al. (2003), involving two Population~III supernovae, predicts an
  oxygen abundance of $\mbox{[O/Fe]}= +4.1$ for {\he}, in strong contradiction
  with the observed value. In conclusion, none of the above mentioned scenarios,
  in their present realizations, can satisfactorly explain the abundance pattern
  of {\he}.
\end{abstract}



\section{INTRODUCTION}

{\he} has been found by \citet[][ hereafter Paper~I and Paper~II,
respectively]{HE0107_Nature,HE0107_ApJ} to be a giant with an iron abundance of
$\mbox{[Fe/H]}_{\mbox{\scriptsize NLTE}} = -5.3$, and large overabundances of
carbon and nitrogen ($\mbox{[C/Fe]}=+4.0$\,dex; $\mbox{[N/Fe]}=+2.3$). These
authors discuss possible origins of the abundance patterns of this
star. Complementary ideas and additional scenarios were presented by
\citet{Shigeyamaetal:2003}, \citet{Bonifacioetal:2003},
\citet{Umeda/Nomoto:2003}, \citet{Schneideretal:2003}, and
\citet{Limongietal:2003}.

Bonifacio et al. emphasized that the oxygen abundance of {\he} would be a
critical observation for constraining possible scenarios. They propose that if
$\mbox{[O/Fe]}>+3.5$, {\he} was formed from a gas cloud which has been
enriched by the yields of at least two supernovae of type II (hereafter SN~II)
of zero metallicity. In contrast, if $\mbox{[O/Fe]}<+3.0$, the high CN
abundances at the surface of {\he} could be caused by internal or external
pollution. In the latter case, {\he} would have a formerly more massive
companion which transferred material to the surface of {\he} during its AGB
phase.

Umeda \& Nomoto proposed that the abundance pattern of {\he} arises from
material that has been enriched by a single $25\,M_{\odot}$ Population~III star
exploding as a supernova with an explosion energy of only $E_{\mbox{\scriptsize
exp}}=3\cdot 10^{50}\,\mbox{erg}$. By assuming that the material produced during
the SN event is homogeneously mixed over a wide range of the mass coordinate,
and a large fraction of the material falls back onto the compact remnant, Umeda
\& Nomoto are able to reproduce the abundance pattern of {\he} remarkably well
(see their Figure 1). Their model predicts that the oxygen abundance of {\he} is
$\mbox{[O/Fe]}\sim 2.8$\,dex.

In this paper we report on new radial velocity measurements for {\he}, and the
analysis of OH lines from newly-obtained ultra-violet VLT/UVES spectra for this
star and {\cd} for comparison. The LTE oxygen abundance was derived by means of
spectrum synthesis. Plane-parallel MARCS model atmospheres were used with
atmospheric parameters 5100\,K, $\log g = 2.20$, $\mbox{[M/H]}=-5.4$ and
$\mbox{[C/Fe]}=+4.0$, $\mbox{[N/Fe]}=+2.3$ for {\he} and 4900\,K, $\log g=2.0$,
and $\mbox{[M/H]}=-4$ for {\cd}. A microturbulence value of 2.0\,km\,s$^{-1}$
was indicated for {\he} and 2.5\,km\,s$^{-1}$ for {\cd}. When deriving CNO and
other metal abundances, the atomic and molecular equilibria and opacities were
recomputed for the different abundances but the temperature structure of the
models were not changed. The newly derived abundances of elements other than O
will be reported in a later paper.

\section{OBSERVATIONS}

New spectra for {\he} and {\cd} were secured in the period 30 September to 27
December 2002 at the European Southern Observatory (ESO), Paranal, Chile, with
the Ultraviolet-Visual Echelle Spectrograph \citep[UVES;][]{Dekkeretal:2000}
mounted on the 8\,m Unit Telescope 2 (Kueyen) of the Very Large Telescope
(VLT). UVES was used in dichroic mode and with various settings. With the BLUE
346\,nm setting covering 3050--3870\,{\AA}, a total of 20\,h were spent on
{\he}, with individual exposures not exceeding 1\,h in order to facilitate
removal of cosmic ray hits.  A $1''$ slit was used in the blue arm, yielding a
resolving power of $R=40,000$. The coadded and rebinned spectra have $S/N\sim
30$ per 0.05\,{\AA} pixel. The spectra of {\cd} have a comparable resolution and
$S/N$.

\begin{figure*}[htbp]
  \epsfig{file=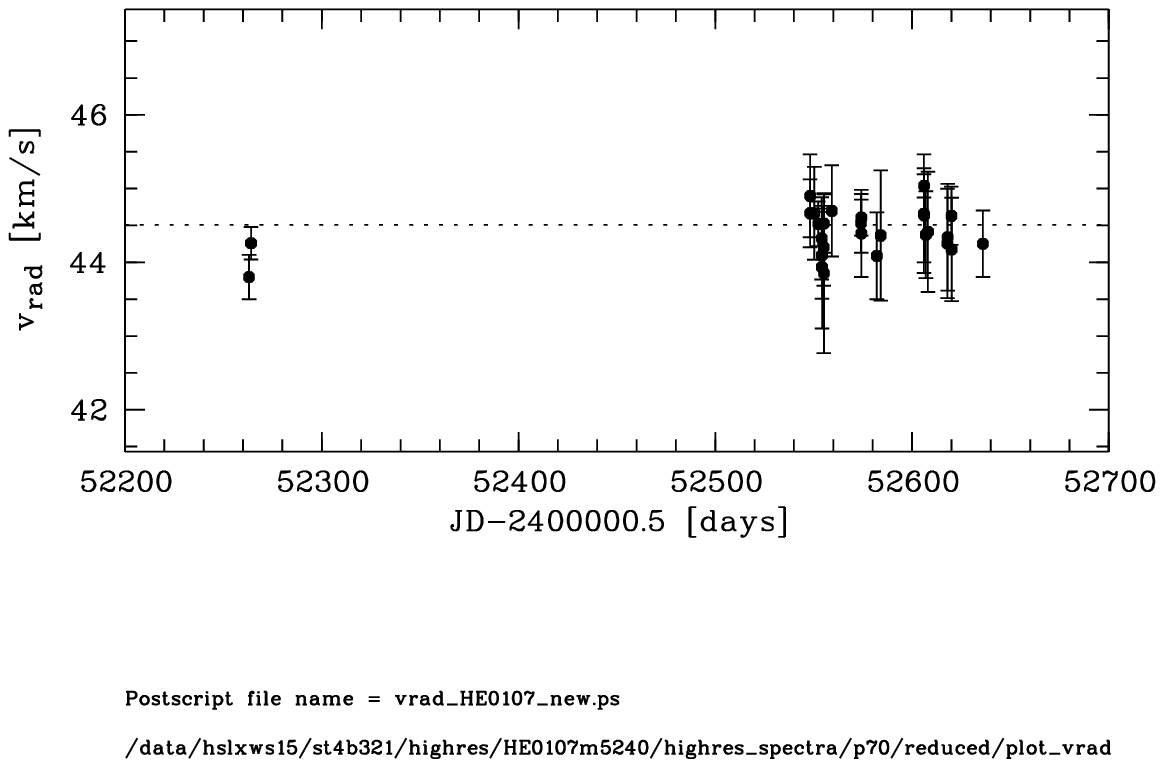, clip=, width=\textwidth,
    bbllx=76, bblly=427, bburx=413, bbury=585}
  \centering
  \caption{\label{Fig:vrad_HE0107_new} Barycentric radial velocities of {\he},
    derived from high-resolution UVES spectra. The error bars indicate the
    $1\,\sigma$ scatter of measurements of individual lines. The dotted line is
    the average of all measurements. They are consistent with a constant radial
    velocity of $44.5$\,km/s during the period covered by the observations (373
    days). }
\end{figure*}

Including these new spectra, radial velocity ({\vrad}) measurements for {\he}
based on high-resolution spectra now cover a period of 373 days. Over this
period the radial velocity was constant at $44.5$\,km/s with a $1\,\sigma$
scatter of the measurements of 0.5\,km/s (see Figure \ref{Fig:vrad_HE0107_new}).

\section{THE LINE LISTS}

Data for atomic and molecular lines for the spectrum synthesis were taken from
several sources. For preference, LIFBASE \citep{Luque/Crosley:1999}, based on a
selection of recent molecular constants, was used for the molecular
lines. LIFBASE was used to generate the wavelengths and transition probabilities
for the various lines associated with different branches for the A-X OH band,
but the corresponding upper and lower energy levels for each line were obtained
by cross-correlating the wavelengths, branches and $J$ values of the transitions
with those in the \citet{Kurucz:1994} molecular database. For $J$ values higher
than provided by Kurucz, energy levels were extrapolated.
 
\section{THE OXYGEN ABUNDANCES}

\begin{figure*}[htbp]
  \epsfig{file=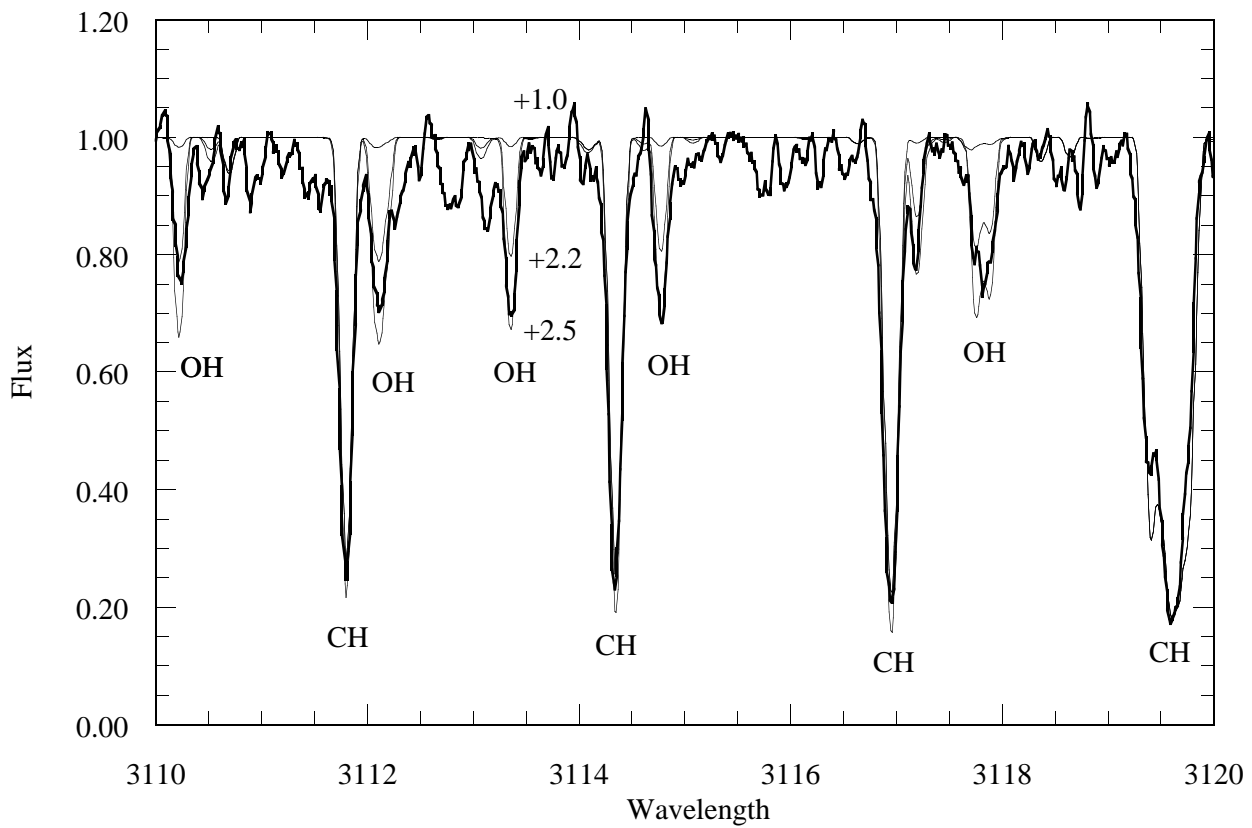, clip=, width=\textwidth,
    bbllx=34, bblly=238, bburx=534, bbury=572}
  \centering
  \caption{\label{Fig:he311-312} Synthetic spectra of {\he} for
    $\mbox{[O/Fe]}=+1.0$, $+2.3$ and $+2.5$ (thin lines). The observed spectrum
    is shown by the thick line. }
\end{figure*}

The median filtered UV spectra for {\he} and {\cd} were flattened by eye over
50\,{\AA} sections and normalized.  In {\he}, the spectrum below 3200\,{\AA} is
dominated by strong lines of the CH C-X band, however where the CH lines thin
out between 3090\,{\AA} and 3120\,{\AA} unblended lines due to OH are clearly
visible. In {\cd} where only the CH band head at 3144\,{\AA} is visible, many
strong unblended OH lines are evident between 3090\,{\AA} and 3200\,{\AA}. In
{\he} we fitted an oxygen abundance of $\mbox{[O/Fe]}=+2.4$ as shown in
Fig. \ref{Fig:he311-312}; for {\cd} we derive $\mbox{[O/Fe]} = +1.0$. We adopted
a solar oxygen abundance of $\log \epsilon(\mbox{O}) = 8.66$\,dex (on a scale
where $\log\epsilon(\mbox{H})=12$\,dex), as determined by
\citet{Asplundetal:2004} from [O\,I], O\,I and OH lines using 3D simulations of
the solar atmosphere, and we adopted a solar iron abundance of $\log
\epsilon(\mbox{Fe}) = 7.45$\,dex \citep{Asplundetal:2000b}. Concerning the Fe
abundance of {\he}, we use our previously determined value of
$\mbox{[Fe~I/H]}_{\mbox{\scriptsize NLTE}} = -5.3$, i.e., $\log
\epsilon(\mbox{Fe}) = 2.2$\,dex (Paper~I and II). We note in passing that this
value agrees to within $0.03$\,dex with the Fe~I abundance measured from 49
lines detected in the newly-obtained UV spectra of {\he} (Bessell et al. 2004,
in preparation).

In order to compare our [O/Fe] value for {\he} with the prediction of Umeda \&
Nomoto, we have to take into account that they use (Nomoto 2003, private
communication) the solar photospheric abundances of
\citet{Anders/Grevesse:1989}, i.e., $\log \epsilon(\mbox{O}) = 8.93$\,dex and
$\log \epsilon(\mbox{Fe}) = 7.51$\,dex. Hence, we need to add $0.21$\,dex
to their value of $\mbox{[O/Fe]}=+2.8$ to place it on our scale.

The most serious error in the oxygen abundance derived for {\he} is most
probably a result of our use of a one-dimensional model
atmosphere. \citet{Asplund/GarciaPerez:2001} discuss the use of OH lines for
abundance analysis of metal-poor stars close to the turn-off point, with
application of 3D hydrodynamical model atmospheres.  They found severely reduced
oxygen abundances (i.e. enhanced OH line strenghts) for the 3D models by up to
$-1$\,dex. The effects increase with increasing effective temperature and with
decreasing metallicity. They reflect the basic fact that in the 3D models the
gas of the upper atmosphere is cooled by adiabatic expansion from ``convective
overshoot'', and that the heating by radiation absorbed by spectral lines is
reduced due to the low metallicity. Along numerous vertical rays through the 3D
model atmosphere, the temperature is lowered and starts the photospheric
temperature rise at greater optical depths and with steeper temperature
gradients. This strongly enhances formation of molecules with relatively low
dissociation energies, such as hydrides. For stars cooler than those
investigated by Asplund \& Garc{\'\i}a P{\'e}rez, somewhat smaller effects are
expected: the convective instability is at greater optical depths, and the
temperature sensitivity of the molecular abundance is reduced.

Guided by preliminary 3D hydrodynamical models for metal-poor giants by Collet
\& Asplund (2003, private communication), we have constructed a number of
artificially cooled plane-parallel models, keeping the temperature constant at
the surface temperature to optical depths ranging from $\tau_{5000} = 0.01$ to
$0.03$, and then increasing $T(\tau)$ gradually to the values of the standard
model. Such models lead to reductions of the oxygen abundance by
$0.3$--$0.4$\,dex, yielding $\mbox{[O/Fe]}=2.0\mbox{--}2.1$ for {\he}. However,
we note that one must await taylored 3D models for {\he} before one may exclude
that the effects are even greater.

\section{DISCUSSION AND CONCLUSIONS}

We have not detected any {\vrad} variations exceeding 0.5\,km/s ($1\,\sigma$
level) over a period of 373 days. On the one hand, it can not yet be ruled out
that {\he} is a very long period and/or low amplitude binary, but on the other
hand, apart from a few exceptions, typical periods and {\vrad} amplitudes of
extremely metal-poor stars with strong carbon-enhancements known to be binaries
are $\mbox{P}=1.5\mbox{--}8.5$ years and $A=5\mbox{--}15$\,km/s
\citep[e.g.][]{Preston/Sneden:2001}. The physical reason for this is that in
closer systems, mass-transfer would occur already before the formerly
more-massive companion reaches the AGB state, and in very wide systems,
mass-transfer will not occur at all \citep[see e.g.][ and references
therein]{Ryan:2003}. It thus does not seem very likely that the large
overabundances of CNO in the atmosphere of {\he} have been produced in a
medium-mass AGB star which later evolved to a white dwarf.

The supernova model of \citet{Umeda/Nomoto:2003} of a $\sim 25\,M_{\odot}$
Population~III star that explodes with low explosion energy
($E_{\mbox{\scriptsize exp}}=3\cdot 10^{50}\,\mbox{erg}$) followed by mixing and
fallback do produce element yields that fit the overall abundance pattern of
{\he} rather well; in particular the high C to N and light to heavy element
ratios. However, the oxygen abundance predicted for {\he} is $\mbox{[O/Fe]}\sim
3.0$\,dex (on our scale), while we measure $\mbox{[O/Fe]}\sim 2.4$\,dex, and the
observed value might need to be reduced to $\mbox{[O/Fe]}=2.0\mbox{--}2.1$\,dex
to take into account 3D effects. In other words, the predicted and observed
value might disagree by as much as 1\,dex. It remains to be studied, however,
whether this disagreement can be reduced by modifying one of the free
parameters of the Umeda \& Nomoto model, e.g., the progenitor mass, and/or the
mass coordinate range in which mixing occurs.

%
%

\citet{Limongietal:2003} propose that the gas cloud from {\he} formed has been
pre-enriched by two supernovae of type II (SN~II) with masses of $\sim
35\,M_{\odot}$ and $\sim 15\,M_{\odot}$, respectively.  They predict an oxygen
abundance of $\mbox{[O/Fe]}= +4.1$ for {\he} (see their Figure 3), in strong
contradiction with the observed value. It remains to be explored if a lower
oxygen abundance can be achieved by combining the yields of two SN~II of
different masses.

As also discussed by \citet{Limongietal:2003}, a self-enrichment explanation for
the CNO abundances of {\he}, based on the effects of a He convective shell that
forms at the core He flash and penetrates into the H-rich envelope, appears to
be ruled out due to several independent arguments, including the high C/N and
carbon isotopic ratio of $^{12}\mbox{C}/^{13}\mbox{C} > 50$ observed in {\he},
as well as evolutionary timescale arguments.

We conclude that the scenarios suggested for the abundance pattern of {\he},
based on one or two supernovae, and possibly in combination with self-enrichment
by the star, are in their present realizations ruled out by observations. Also,
the lack of detections of variations in radial velocity makes pollution by a
companion increasingly improbable. It remains to be seen if the scenarios can be
modified to fit the observations better, or whether other ideas must be invented
and explored.

\acknowledgments

We would like to express our gratitude to the ESO staff on Paranal and Garching
for carrying out the observations with VLT-UT2, and reducing the data,
respectively. Thanks are due to Martin Asplund and Remo Collet for discussion on
3D models, and to Bengt Edvardsson for computational assistance. M.S.B. and
N.C. acknowledge support through a \emph{Linkage International Fellowship} of
the Australian Research Council, making possible a 6 months stay of N.C. at
Mt. Stromlo Observatory, where part of the work described in this paper has been
carried out. B.G. thanks the Swedish Research Council for financial support.

\bibliography{atomdata,instr,ncpublications,mphs}
\bibliographystyle{apj}

\clearpage


\end{document}